\documentclass{webofc}
\usepackage[varg]{txfonts}   
\usepackage{hyperref}
\usepackage{cleveref}

\newcommand{\Nmu}{$N_\mu$\xspace}

\newcommand{\Xmax}{$X_\mathrm{max}$\xspace}

\newcommand{\epos}{\textsc{EPOS}\xspace}
\newcommand{\eposlhc}{\textsc{EPOS-LHC}\xspace}
\newcommand{\sibyll}[1]{\textsc{Sibyll#1}\xspace}
\newcommand{\qgsii}{\textsc{QGSJet-II.04}\xspace}
\newcommand{\qgs}{\textsc{QGSJet}\xspace}

\begin{document}
\title{Probing Hadronic Interactions with Cosmic Rays}
%
%

\author{\firstname{Dennis} \lastname{Soldin}\inst{1,2}\fnsep\thanks{\email{soldin@kit.edu}} }

\institute{Karlsruhe Institute of Technology, 76021 Karlsruhe, Germany
\and University of Delaware, Newark, DE 19716, USA
}

\abstract{%
High-energy cosmic rays interact in the Earth's atmosphere and produce extensive air showers (EASs) which can be measured with large detector arrays at the ground. The interpretation of these measurements relies on models of the EAS development which represents a challenge as well as an opportunity to test quantum chromodynamics (QCD) under extreme conditions. The EAS development is driven by hadron-ion collisions under low momentum transfer in the non-perturbative regime of QCD. Under these conditions, hadron production cannot be described using first principles and these interactions cannot be probed with existing collider experiments. Thus, accurate measurements of the EAS development provide a unique probe of multi-particle production in hadronic interactions. 
}
\maketitle
\section{Introduction}
\label{intro}
Throughout the history of elementary particle physics, discoveries have been made through the observation of cosmic rays and neutrinos. Examples are the discovery of new elementary particles, the confirmation of neutrino oscillations, as well as measurements of particle interactions far beyond current collider energies. When a cosmic ray enters the atmosphere and collides with an air nucleus, it initiates particle cascades which form an extensive air shower (EAS). While the decay of neutral pions into photon pairs produces an electromagnetic cascade, charged pions, kaons, and baryons interact again with air nuclei. This process continues until most energy is dissipated through the electromagnetic cascade, and charged particles reach an energy where the decay into muons becomes more likely than (re-)interactions with air nuclei. Thus, muons are generally messengers of the hadronic interactions in EASs.


Experimentally, the properties of the initial cosmic ray, such as energy and mass, are inferred indirectly from the particles measured at the ground by large detector arrays and their interpretation strongly relies on simulations of the EAS development. The main challenge in the description of EASs is the treatment of hadronic interactions in the atmosphere over many orders of magnitude in energy. While interactions of hadrons up to several TeV have been well studied at collider experiments, interactions at higher energies can not be probed by existing facilities. Interactions in the forward direction, which dominate the particle production in EASs, and the combination of particles involved in those interactions can not be measured with current detectors at the Large Hadron Collider (LHC) at CERN. The EAS development is mainly driven by relativistic hadron-ion collisions in the atmosphere at low momentum transfer in the non-perturbative regime of quantum chromodynamics (QCD). Under these conditions, hadron production cannot be described using first principles and thus existing simulation packages rely on a variety of phenomenological hadronic interaction models. Commonly used hadronic interaction models are \sibyll{}~\cite{Ahn:2009wx,Riehn:2019jet}, \qgs~\cite{Ostapchenko:2005nj, Ostapchenko:2019few}, \epos~\cite{Werner:2005jf, Pierog:2013ria}, and \textsc{DPMJet}~\cite{Ranft:1994fd}. Generally, all these models are based on different realizations of perturbative QCD with Gribov-Regge theory and rely on some fundamental principles, such as conservation laws. However, the particle production is dominated by non-perturbative QCD which is treated by more phenomenological approaches (see e.g., Ref.~\cite{Albrecht:2021cxw} for a recent review). Large uncertainties remain due to those theoretical limitations and the lack of data from existing collider experiments. Thus, dedicated measurements of hadronic interactions in EASs are crucial in order to test and improve existing hadronic interaction models.

\section{Hadronic Interactions and the Muon Puzzle in EASs}
\label{sec:hadronic-interactions}

\begin{figure}[b]
    \centering
    \vspace{-1em}
    \mbox{\hspace{-0.5em}%
    \includegraphics[width=0.45\textwidth]{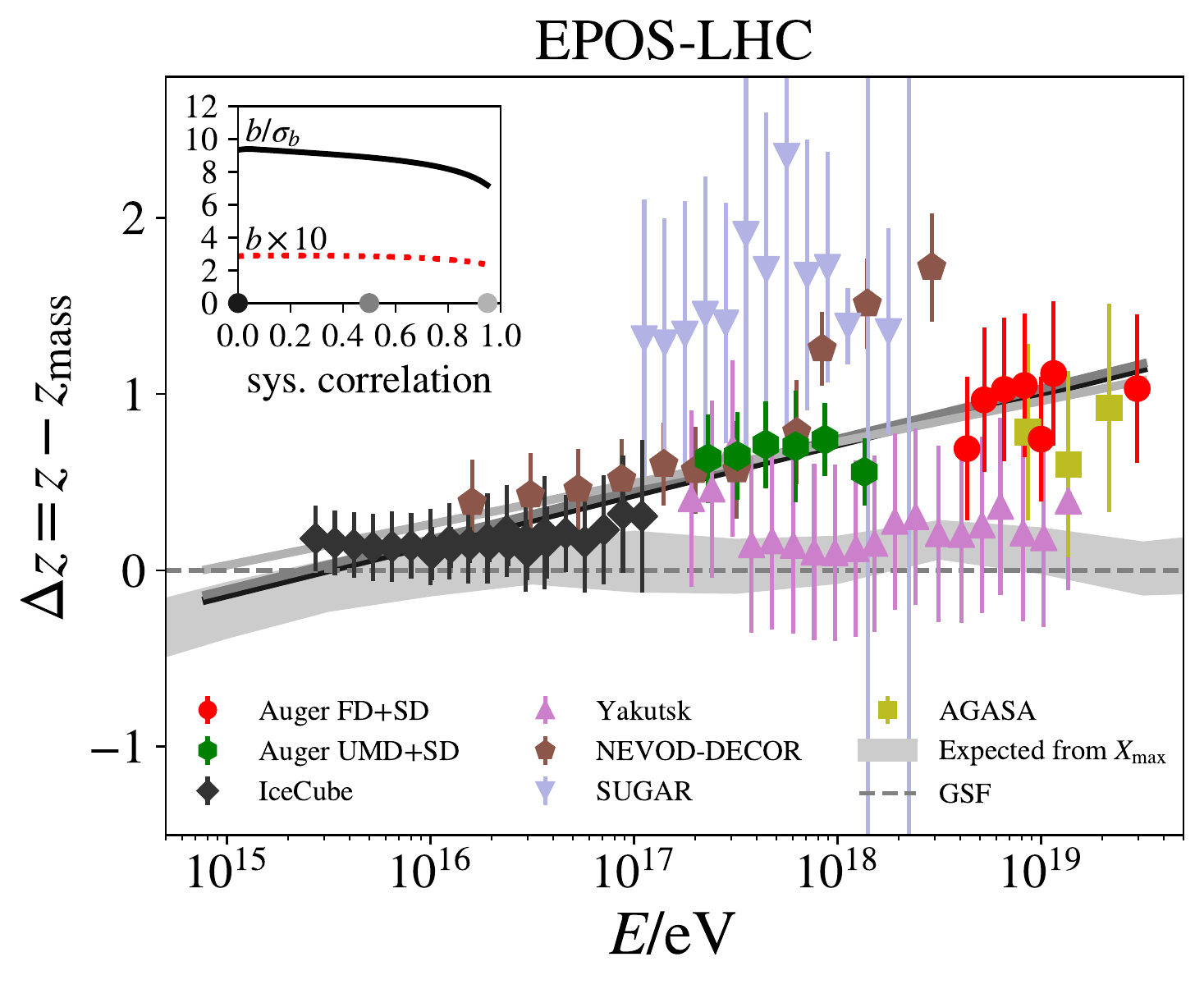}\;\;\;
    \includegraphics[width=0.45\textwidth]{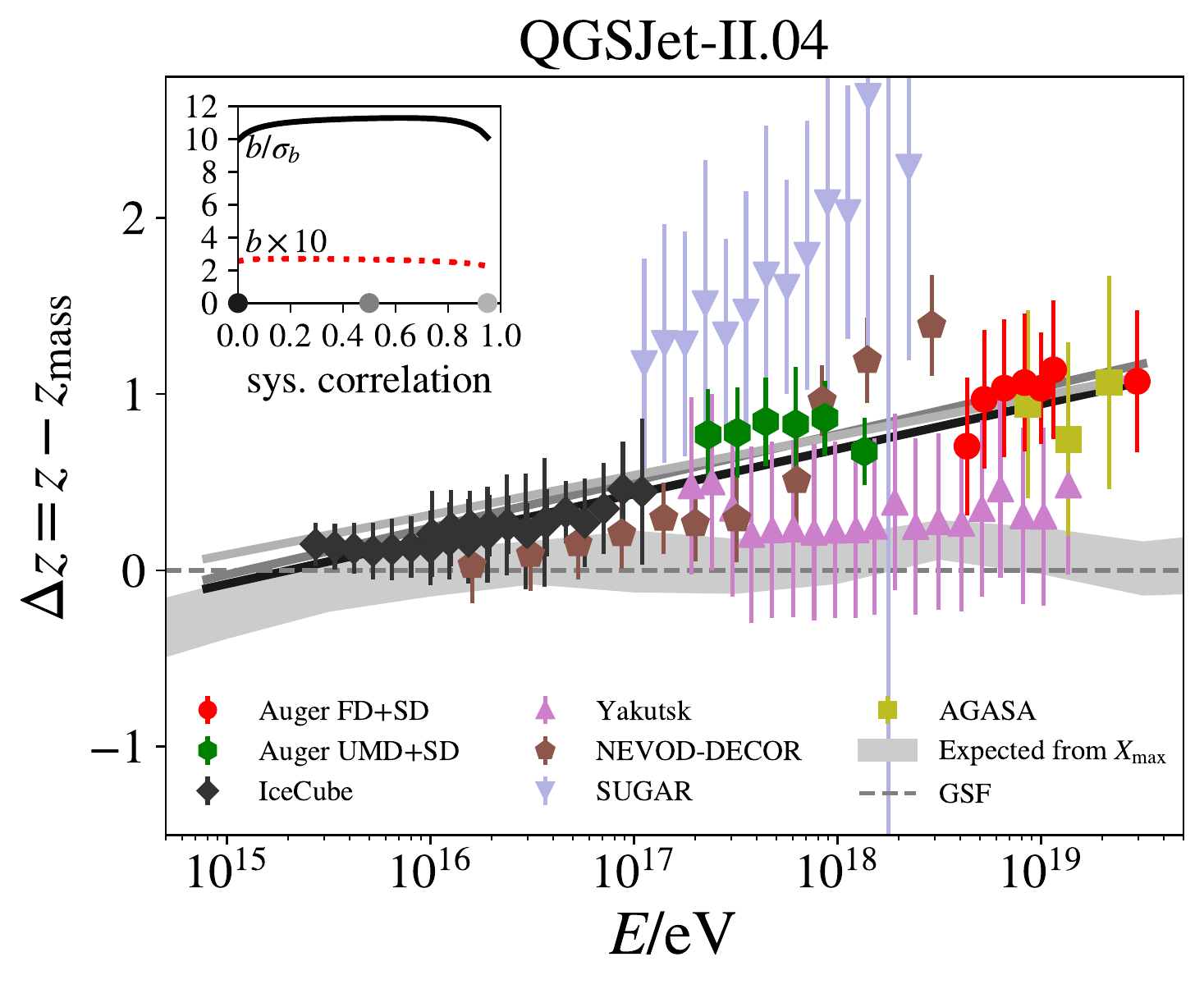}%
    }
    \vspace{-.8em}
    
    \caption{Linear fits of the form $\Delta z=a+b\cdot \log_{10}(E/10^{16}\,\mathrm{eV})$ to the $\Delta z=z-z_\mathrm{mass}$ distributions, as described in Ref.~\cite{EAS-MSU:2019kmv}. Shown in the inset are the slope, $b$, and its deviation from zero in standard deviations for an assumed correlation of the uncertainties within each experiment (for details see Ref.~\cite{Soldin:2021wyv}). Examples of the fits are shown for a correlation of $0.0$, $0.5$, and $0.95$ in varying shades of gray.} 
    \label{fig:WHISP}%
    \vspace{-1.8em}
    
\end{figure}

Muons are tracers of hadronic interactions and thus the measurement of the number of muons in EASs, \Nmu, is an important observable in order to test hadronic interaction models. Over the last 20 years, many experiments reported discrepancies between model predictions and experimental data, while other experiments reported no discrepancies~\cite{Albrecht:2021cxw}. 
A systematic meta-analysis of measurements of GeV muons from nine air shower experiments 
found a global picture of the energy dependence and high statistical significance of the discrepancies~\cite{EAS-MSU:2019kmv,Soldin:2021wyv}. Key aspects of this analysis are a cross-calibration of the energy scales of the different experiments and the definition of the $z$-value, an abstract measure of the muon content which is comparable between experiments and different analyses. The value $\Delta z=z-z_\mathrm{mass}$, where $z_\mathrm{mass}$ is the number of muons predicted by a hadronic model assuming a mass composition of the primaries based on experimental parameterizations~\cite{Dembinski:2017zsh}, measures the difference between the experimental data and the inferred number of muons for a given hadronic interaction model. A positive value indicates an excess of muons in data with respect to simulations and $0$ indicates a perfect match. The resulting distribution of $\Delta z$ is shown in Fig.~\ref{fig:WHISP} for two hadronic models, \eposlhc~\cite{Pierog:2013ria} and \qgsii~\cite{Ostapchenko:2019few}. The experimental measurements are consistent within uncertainties with predictions up to EAS energies of approximately $100$\,PeV. 
However, above $100$\,PeV a muon excess is observed that systematically increases with the EAS energy. The slope of a linear fit to this excess is found to be non-zero at $\sim 8\sigma$ for all hadronic interaction models considered. These discrepancies are referred to as the Muon Puzzle and they indicate severe shortcomings in the understanding of particle physics.

Another important observable to probe hadronic interactions is the standard deviation of the muon number, $\sigma(N_\mu)$, which shows 
reasonable agreement between experimental data~\cite{PhysRevLett.126.152002} and the models \eposlhc~\cite{Pierog:2013ria}, \qgsii~\cite{Ostapchenko:2019few}, and \sibyll{\,2.3d}~\cite{Riehn:2019jet}. 
In addition, discrepancies in measurements of the attenuation length of GeV muons~\cite{Apel:2017thr} suggest a problem in the description of the energy spectra of muons at production. Preliminary studies  also indicate differences in the discrepancies of GeV and TeV muons~\cite{IceCube:2021ixw} and inconsistencies in the zenith angle distributions of simulated TeV muons have been reported~\cite{Soldin:2018vak}. 
The muon production depth in the atmosphere~\cite{Apel:2011zz,PhysRevD.90.012012} has also been shown to have discrepancies with respect to model predictions, which has been related to a poor description of pion-air interactions in EASs. These observations combined provide strong constraints on changes of certain model parameters that have a large impact on the number of muons in EAS simulations.

\section{Constraining Hadronic Interaction Models}
\label{sec:accelerators}

\begin{figure}[b]
\centering
\vspace{-1em}

\includegraphics[width=0.85\textwidth]{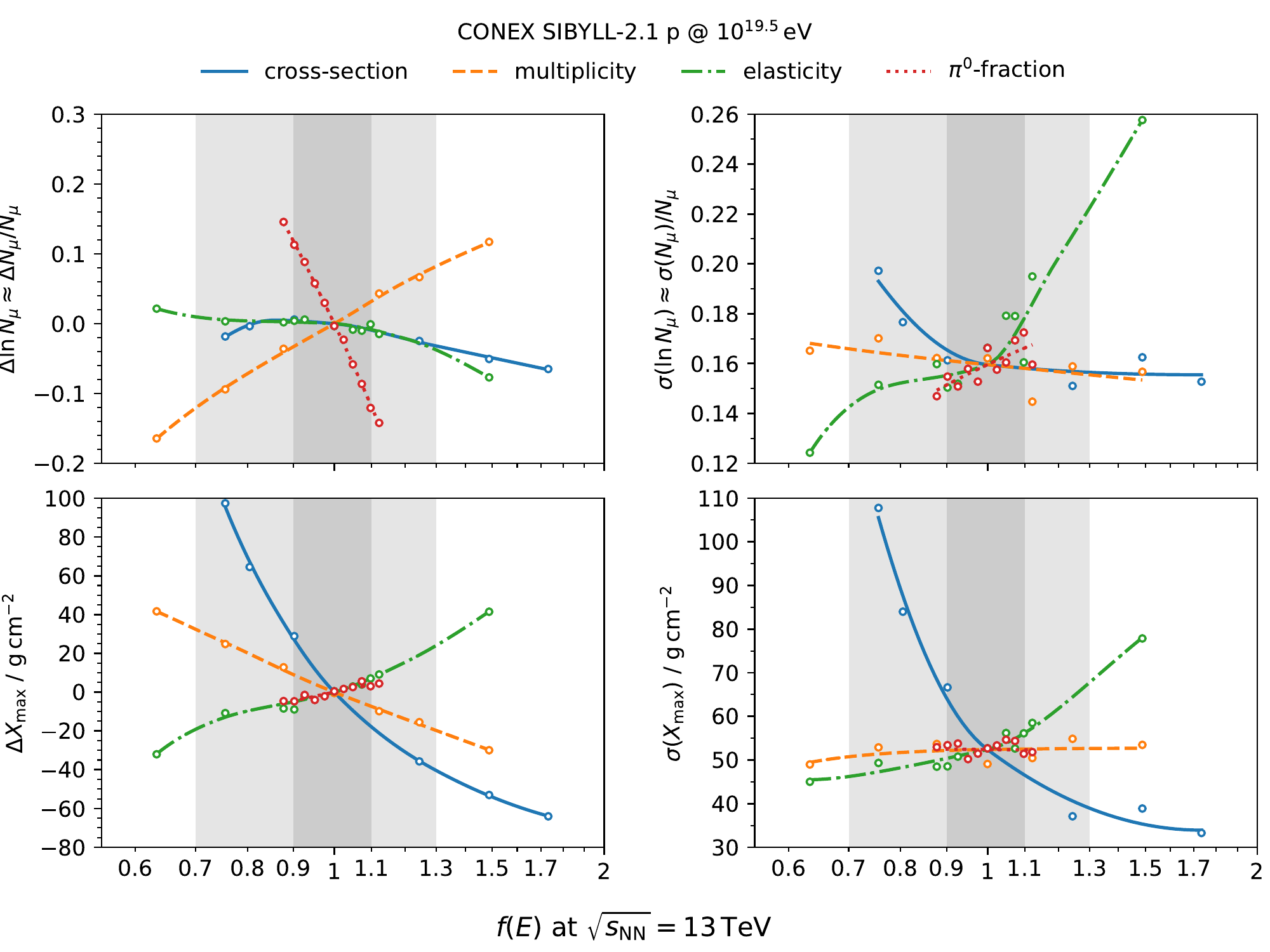}
\vspace{-.8em}
    \caption{Impact of changing basic parameters of hadronic interactions at $\sqrt{s_\mathrm{NN}} = 13\,\mathrm{TeV}$ and extrapolating logarithmically (see Ref.~\cite{Ulrich:2010rg} for details) on the means and standard deviations of the logarithm of the muon number \Nmu (top row) and the depth \Xmax of the shower maximum (bottom row) for a $10^{19.5}\,\mathrm{eV}$ proton shower simulated with \sibyll{\,2.1}~\cite{Ahn:2009wx}. 
The shaded bands highlight a $\pm 10\,\%$ and $\pm 30\,\%$ modification. The figure is an update of the original data from Ref.~\cite{Ulrich:2010rg}, taken from Ref.~\cite{Albrecht:2021cxw}.}
\label{fig:impact_study}
\vspace{-1.em}

\end{figure}

Four basic parameters of hadronic interactions are most relevant for the EAS development: the inelastic cross section for hadrons in air, the hadron multiplicity, the elasticity (the energy fraction carried by the most energetic particle), and the ratio of electromagnetic to hadronic energy flow. The impact of modifications to these aspects on air shower observables has been investigated in Ref.~\cite{Ulrich:2010rg}. The results for the mean and standard deviation of the muon number, \Nmu, and the depth of the shower maximum, \Xmax, for a $10^{19.5}$\,eV proton shower, are shown in Fig.~\ref{fig:impact_study}. The baseline prediction of the \sibyll{\,2.1} model \cite{Ahn:2009wx} was modified with a factor that is $1$ for an energy of $10^{15}\,\mathrm{eV}$ and increases or decreases logarithmically with the beam energy, depending on the value $f(E)$ at some intermediate scale (here $\sqrt{s_\mathrm{NN}} = 13\,\mathrm{TeV}$). For a thorough description of $f(E)$ and a detailed discussion of the modifications see Refs.~\cite{Albrecht:2021cxw, Ulrich:2010rg}.

As shown in Fig.~\ref{fig:impact_study}, the most effective way to increase \Nmu in EASs is to decrease the $\pi^0$-fraction. The number of muons also increases with the multiplicity, but the effect is much weaker. In addition, the observations of the standard deviation of \Nmu~\cite{PhysRevLett.126.152002} provides strong constraints on changes to the elasticity, which has a large impact on the muon fluctuations. Since air shower simulations give a reasonable description of the fluctuations, $\sigma(N_\mu)$, and the depth of the shower maximum, \Xmax, it is important to minimize the impact of changes on $\sigma(N_\mu)$, as well as \Xmax and its standard deviation. This indicates that the only plausible way to increase the muon number sufficiently is to decrease the energy fraction lost to photon production (mainly from $\pi^0$-decay) in hadron collisions. There are multiple possibilities to realize a reduced electromagnetic energy fraction within different models which are reviewed in detail in Ref.~\cite{Albrecht:2021cxw}, for example.


\section{Conclusions and Outlook}
\label{sec:conclusions}

Over the last 20 years, discrepancies between EAS simulations and experimental data have been reported. A first combined meta-analysis of muon data from multiple experiments confirmed an excess of atmospheric muons in EASs above $100$\,PeV. 
Dedicated studies of the impact of key parameters in hadronic interaction models indicate that the only plausible way to increase the muon number is to decrease the energy fraction lost to photon production in hadron collisions. However, in order to confirm these studies, further experimental tests are required.
Updates of various analyses from existing and new air-shower facilities~\cite{Coleman:2022abf}, which include measurements of the muon production depth, the muon energy deposit, and measurements of muons at multiple energies, as well as studies of shower-to-shower fluctuations, will provide crucial information on hadron interactions beyond the phase space of existing colliders to constrain muon production models. In addition, dedicated collider measurements will provide complementary information of hadron production in the forward region~\cite{Albrecht:2021cxw,Feng:2023}. 

%

\bibliography{references.bib}
%
%
%
%

\end{document}